



\magnification\magstep1
\hsize=6 truein
\vsize=8.5 truein
\hoffset=.2 truein
\voffset=0 truept

\parskip=0pt plus .5pt

\raggedbottom

\tolerance=4000


\newcount\eqautonumber
\eqautonumber=0

\def\eqna{\global\advance\eqautonumber by 1\eqno(\the\eqautonumber)}


\newcount\sectnum
\sectnum=0

\outer\def\numsection#1\par{\vskip8truemm plus 4 truemm minus 1 truemm
\advance\sectnum by 1
{\bf
\rightskip=0em plus 12 em
\parindent=1.5pc
\item{\hbox to \parindent{\bf\enspace\the\sectnum.\hfil}}#1\par}
\nobreak\vskip4 truemm
plus 1 truemm minus 1 truemm
\message{\the\sectnum. #1}
\noindent}

\outer\def\nonumsection#1\par{\vskip8truemm plus 4 truemm minus 1 truemm
{\bf
\rightskip=0em plus 12 em
\noindent#1\par}
\nobreak\vskip4 truemm
plus 1 truemm minus 1 truemm
\message{#1}
\noindent}


\newcount\refno
\refno=0
\def\nref#1\par{\advance\refno by1
  \vskip1pt
  \item{\the\refno.\thinspace}#1}

\def\totalno{\message
  {***Total:\the\refno  references***}}


\def\book#1[[#2]]{{\it#1\/} (#2)}

\def\prd#1 #2 #3.{{\it
  Phys.\ Rev.\ \rm D\bf#1} (#2) #3}
\def\plb#1 #2 #3.{{\it
  Phys.\ Lett.\ \bf#1\/}B (#2) #3}
\def\npb#1 #2 #3.{{\it
  Nucl.\ Phys.\ \rm B\bf#1} (#2) #3}
\def\jmp#1 #2 #3.{{\it
  J.\ Math.\ Phys.\ \bf#1} (#2) #3}
\def\cqg#1 #2 #3.{{\it
  Class.\ Quantum Grav.\ \bf#1} (#2) #3}
\def\grg#1 #2 #3.{{\it
  Gen.\ Rel.\ Grav.\ \bf#1} (#2) #3}
\def\cmp#1 #2 #3.{{\it
  Commun.\ Math.\ Phys.\ \bf#1} (#2) #3}
\def\phrep#1 #2 #3.{{\it
  Phys.\ Rep.\ \bf#1} (#2) #3}
\def\annph#1 #2 #3.{{\it
  Ann.\ Phys.\ }(N.\thinspace Y.)
  {\bf#1} (#2) #3}
\def\annphg#1 #2 #3.{{\it
  Ann.\ Phys.\ }(Germany)
  {\bf#1} (#2) #3}
\def\annmath#1 #2 #3.{{\it
  Ann.\ Math.\ }
  {\bf#1} (#2) #3}
\def\prl#1 #2 #3.{{\it
  Phys.\ Rev.\ Lett.\ \bf#1} (#2) #3}
\def\pr#1 #2 #3.{{\it
  Phys.\ Rev.\ \bf#1} (#2) #3}
\def\mpla#1 #2 #3.{{\it
  Mod.\ Phys.\ Lett.\ \rm A\bf#1} (#2) #3}
\def\ijmpa#1 #2 #3.{{\it
  Int.\ J.~Mod.\ Phys.\ \rm A\bf#1} (#2) #3}
\def\ncimb#1 #2 #3.{{\it
  Nuovo Cim.\ \bf#1\/}B (#2) #3}
\def\ncim#1 #2 #3.{{\it
  Nuovo Cim.\ \bf#1\/} (#2) #3}
\def\advphys#1 #2 #3.{{\it
  Adv.\ Phys.\ \bf#1} (#2) #3}
\def\jphysa#1 #2 #3.{{\it
  J.\ Phys.\ \rm A\bf#1} (#2) #3}
\def\foundph#1 #2 #3.{{\it
  Found.\ Phys.\ \bf#1} (#2) #3}
\def\sovpu#1 #2 #3.{{\it
  Sov.\ Phys.\ Usp.\ \bf#1} (#2) #3}
\def\sovpjetp#1 #2 #3.{{\it
  Sov.\ Phys.\ JETP \bf#1} (#2) #3}
\def\zetf#1 #2 #3.{{\it
  Zh.\ Eksp.\ Teor.\ Fiz.\ \bf#1} (#2) #3}
\def\repprogph#1 #2 #3.{{\it
  Rep.\ Prog.\ Phys.\ \bf#1} (#2) #3}
\def\phscripta#1 #2 #3.{{\it
  Phys.\ Scripta \bf#1} (#2) #3}
\def\aipoincare#1 #2 #3.{{\it
  Ann.\ Inst.\ Henri Poincar\'e \bf#1} (#2) #3}
\def\rmp#1 #2 #3.{{\it
  Rev.\ Mod.\ Phys.\ \bf#1} (#2) #3}
\def\prsa#1 #2 #3.{{\it
  Proc.\ R.\ Soc.\ London \rm A\bf#1} (#2) #3}


\font\titlefont=cmbx12
\font\abstractfont=cmr10 at 10 truept
\font\authorfont=cmcsc10
\font\addressfont=cmsl10 at 10 truept
\font\smallbf=cmbx10 at 10 truept

\font\smallrm=cmr7


\def\tref#1/{$^{#1}$}

\def\half{{1\over2}}
\def\psnb{\Psi_{\hbox{\smallrm NB}}}


\hyphenation{
mini-su-per-space
pre-factor
an-iso-tropic
mani-fold
mani-folds
}



\line{\hfill $\hbox to 5 truecm{\hfil SU-GP-93/5-3}
\atop
\hbox to 5 truecm{\hfil gr-qc/9305016}$}

\hskip 1.5 truecm

{\baselineskip=24 truept
\titlefont
\centerline{INITIAL CONDITIONS AND UNITARITY IN}
\centerline{UNIMODULAR QUANTUM COSMOLOGY}
}
\footnote{}{\abstractfont Talk given at the 5th Canadian Conference on
General Relativity  and Relativistic Astrophysics,
Waterloo, Ontario, Canada,
May 1993.}

\vskip .7 truecm plus .3 truecm minus .2 truecm

\centerline{\authorfont Alan Daughton, Jorma Louko and Rafael D. Sorkin}
\vskip 2 truemm
{\baselineskip=12truept
\addressfont
\centerline{Department of Physics, Syracuse University}
\centerline{Syracuse, NY 13244--1130, USA}
}

\vskip 1.1 truecm plus .3 truecm minus .2 truecm

\centerline{\smallbf Abstract}
\vskip 1 truemm
{\baselineskip=12truept
\leftskip=3truepc
\rightskip=3truepc
\parindent=1truepc

\abstractfont

We consider quantization of the positive curvature Friedmann cosmology
in the unimodular modification of Einstein's theory, in which the spacetime
four-volume appears as an explicit time variable. The Hamiltonian admits
self-adjoint extensions that give unitary evolution in the Hilbert space
associated with the Schr\"odinger equation. The semiclassical estimate to
the no-boundary wave function of Hartle and Hawking is found. If this
estimate is accurate, there is a continuous flux of probability into the
configuration space from vanishing three-volume, and the no-boundary wave
function evolves nonunitarily. Generalizations of these results hold in a
class of anisotropic cosmologies.
\par}

\vskip .3 truecm plus .3 truecm minus .2 truecm

\nonumsection

In formulating the gravitational functional integral
$\int {\cal D} g\, \exp \left[iS(g)\right]$, one may choose to limit the
geometries $g$ which enter the sum by specifying---in addition to any
boundary conditions which may be imposed on $g$---a fixed value $T$
for the total four-volume: $\int \sqrt{-g} d^4 x = T$.  This
produces a theory\tref1-8/ whose classical limit is equivalent to the
Einstein theory except that the cosmological constant becomes a constant of
integration, rather than a dynamically unalterable parameter in the
Lagrangian.  Limiting $g$ in this way (which is particularly natural in the
context of causal set theory) not only tends to ameliorate the convergence
difficulties of the functional integral, but it does away with the
``frozen'' character of the wave function~$\Psi$,  replacing one of the
Hamiltonian constraint equations with a Schr\"odinger equation for the
{\it evolution\/} of $\Psi$ with~$T$.  (In addition to its technical
ramifications, this ``unfreezing'' helps in making physical
sense of questions like whether quantum effects circumvent the
singularities of gravitational collapse.)

In both the sum-over-histories interpretation of this ``unimodular''
theory\tref1/ (in which the history itself is the primary object) and the
Hilbert space interpretation\tref2-4/ (in which the wave function is the
primary object) it is important to ask whether $\Psi$ evolves unitarily
with~$T$.  In this contribution we shall study the wave function derived
from the no-boundary proposal of Hartle and Hawking,\tref9-11/ whose aim is
in effect to specify initial boundary conditions on~$g$.  Specifically, we
ask: does the no-boundary prescription give a wave function which evolves
unitarily in the Hilbert space of the unimodular theory?

Recall first that the no-boundary proposal of Hartle and Hawking is a
topological statement about the manifolds that are taken to contribute to
the path integral in terms of which the wave function is defined. In
conventional Einstein gravity one writes\tref12/
$$
\psnb
\left( h_{ij}; \Sigma \right) =
\int
{\cal D} g_{\mu\nu} \,
\exp \left[ - I \left( g_{\mu\nu};  M \right) \right]
\ \ ,
\eqna
$$
where $I \left( g_{\mu\nu}; M \right)$ is the Euclidean action of
the gravitational field $g_{\mu\nu}$ on the four-manifold~$M$. The
four-manifold $M$ is compact with a boundary, such that its boundary is the
three-surface $\Sigma$ which appears in the argument of the wave function.
The integral is over metrics $g_{\mu\nu}$ on $M$ which induce the
metric $h_{ij}$ on~$\Sigma$. To give a meaning to this formal expression,
additional input is needed; see for example Refs.\tref12,13/

To apply these ideas in unimodular gravity, we take the unimodular
no-boundary wave function to be given by
$$
\psnb
\left( h_{ij}; \Sigma; T \right) =
\int
{\cal D} g_{\mu\nu} \,
\exp \left[ iS \left( g_{\mu\nu}; M \right) \right]
\ \ ,
\eqna
$$
where $M$ and $\Sigma$ are as above, and $S$ is the Lorentzian action.
The conditions for the metrics $g_{\mu\nu}$ are that they induce the
metric $h_{ij}$ on~$\Sigma$, and that the total Lorentzian four-volume is
equal to~$T$. Note that, even apart from issues of convergence, the
explicit appearance of the {\it Lorentzian\/} four-volume in the
conditions for $g_{\mu\nu}$ implies that the integral (2) can in no sense be
thought of as an integral over Euclidean (i.e., positive definite) metrics.

As an example we consider the positive curvature Friedmann model. The
metric is
$$
ds^2 = \sigma^2
\left[ - N^2 (t) dt^2 + a^2 (t) d \Omega_3^2 \right]
\ \ ,
\eqna
$$
where $d\Omega_3^2$ is the metric on the unit three-sphere,
$\sigma^2=2G/3\pi$, and $G$ is Newton's constant. We take $a>0$. The
action of the unimodular theory for this metric is obtained by inserting
the metric into the Einstein action and setting $N=1/(a^3)$, with the
result
$$
S = \half \int
\left(- a^4 {\dot a}^2 + {1 \over a^2} \right) \, dt
\ \ .
\eqna
$$
The Lorentzian solutions obtained by varying (4) are the one-parameter
family of de~Sitter spaces, written in the proper-time gauge as
$$
ds^2 =
\sigma^2 \left[
- d\tau^2 + {1 \over \lambda}
\cosh^2 \left( \sqrt{\lambda} \tau \right)
d \Omega_3^2
\right]
\ \ ,
\eqna
$$
where the integration constant $\lambda>0$ is related to the conventional
cosmological constant $\Lambda$ by $\lambda=\sigma^2\Lambda/3$.

The quantum Hamiltonian operator is
$$
{\hat H} = \half \left[
\left( {1\over a^2} {\partial \over \partial a} \right)
\left( {1\over a^2} {\partial \over \partial a} \right)
- {1 \over a^2}
\right]
\eqna
$$
where we have chosen the standard ``covariant" factor ordering. The quantity
$T$ in the general formalism outlined above is now simply replaced by~$t$,
and the $t$-evolution of $\psi$ is given by the Schr\"odinger equation
${\hat H}\psi=i{\partial\psi}/{\partial t}$.

The Hamiltonian (6) is clearly Hermitian in the corresponding inner product
$$
\left(\chi,\psi\right) =
\int_0^\infty
a^2 da \, {\overline \chi} \psi
\ \ .
\eqna
$$
For the Schr\"odinger evolution to be unitary, one needs a Hamiltonian which
is not only Hermitian but self-adjoint.\tref14/ It is straightforward to
show that our Hamiltonian has a one-parameter family of self-adjoint
extensions, the defining domain of each extension being functions
satisfying the boundary condition
$$
\lim_{a\to0} \left(\alpha \psi + {\beta\over a^2}
{\partial \psi \over \partial a} \right)
=0
\eqna
$$
for given real numbers $\alpha,\beta$. The situation is very much like for
the free particle on the half-line, as can be seen by going\tref1/ to the
natural ``position" coordinate $x=a^3$; the potential term in the
Hamiltonian is singular as $x\to0$, but this singularity is so weak
that it does not essentially affect the self-adjointness analysis.

Thus, with the one-parameter family of different self-adjoint extensions of
the Hamiltonian, we obtain a one-parameter family of distinct quantum
theories. In each of these theories the Schr\"odinger equation gives unitary
dynamics in the Hilbert space determined by the inner product~(7).

We now turn to the no-boundary wave function. Following Hartle and
Hawking,\tref10/ we take the four-manifold to be the closed four-dimensional
ball~${\bar B}^4$. The semiclassical estimate to the path integral is
$$
\psnb
( a, t )
\sim
P \exp \left[iS_c
( a, t ) \right]
\eqna
$$
where $P$ is assumed to be slowly varying compared with the exponential
factor, and $S_c( a, t )$ is the action of the classical solution with the
appropriate boundary data. The classical solution is straightforward to find:
it is a globally regular, genuinely complex metric, satisfying the Einstein
equations with a complex-valued cosmological constant, but having by
construction a real Lorentzian four-volume. The action is given by
$$
S_c( a, t ) =
{ia^2 \over 6}
\left\{
5 -
{2 \left[ 1 + (12it/a^4) \right]
\over
\sqrt{1 + (12it/a^4)} - 1}
\right\}
\ \ .
\eqna
$$
The ambiguity in choosing the sign of $\sqrt{-g}$ for a complex-valued
metric\tref12/ induces an ambiguity in the imaginary part of $S_c( a, t )$.
In~(10), we have fixed this ambiguity by the requirement that the path
integral give a well-defined vacuum state for the quantum field theory of
scalar field perturbations on the gravitational background. For a
discussion of this issue in conventional Einstein theory, see
Refs.\tref12,13/

Consider first a semiclassical interpretation of the
result~(10).\tref12/  For $a^4\ll t$, $S_c$ is genuinely complex, and
$\psnb$ does not correspond to classical Lorentzian spacetimes. For $a^4\gg
t$, on the other hand, $S_c$ is almost real and $\psnb$ is rapidly
oscillating. Integrating the equation $p=(\partial S_c /\partial a)$ in this
domain gives a family of approximately Lorentzian spacetimes in which $t$ is
proportional to~$a^3$, and which can be seen to coincide approximately
with the exact Lorentzian solution~(5).  All these Lorentzian
spacetimes can be thought of as describing four-geometries which evolve
classically after they pass through the region~$a^4\approx t$.

Consider then the compatibility of $\psnb$ with  unitary evolution. One
might expect this question to be difficult on the grounds that we do not
have at hand an exact wave function but only a semiclassical estimate.
However, if the estimate (9) is accurate in the sense that the prefactor $P$
is a slowly varying function of $a$ and $t$ compared with the exponential
factor, it is easy to see that $\psnb$ cannot evolve unitarily according to
any of the self-adjoint extensions of~$\hat H$. The reason is that for
$t\to\infty$, (9) gives uniformly in any finite interval in $a$ the estimate
$\psnb(a,t)\sim  \exp \left(\sqrt{4it/3}\right)$, which grows exponentially
in~$\sqrt{t}$. Therefore, if the norm (7) of $\psnb$ is finite, it cannot be
conserved in time, and the evolution of $\psnb$ must be nonunitary.

The interpretation of this failure of unitarity is seen by noticing that
the probability current associated with $\psnb$ is always towards
increasing~$a$. This means that the exponentially increasing probability is
generated by injecting into the configuration space a continuously
increasing probability flux from the boundary $a=0$. In particular,
the self-adjointness boundary conditions (8) must be violated.  If the
probability flux were directed {\it outward\/} through $a=0$, the natural
interpretation would be in terms of recollapse and disappearance of
the universe.  As it is, one might refer to a ``continuous creation of
universes from nothing." It is intriguing that this effect appears to
be rather similar to the ideas of Linde\tref15/ and
Vilenkin\tref16,17/ for specifying a wave function in conventional
Einstein gravity.  [If taken literally, this continuous creation would
seem to imply a ``second quantization of universes'', but perhaps a
more appropriate image would be of a single connected structure in
which multiple spacetimes emerge (with increasing rapidity but decreasing
overall rate of growth)  from a single ``root'', such as an initial,
non-manifold region of a causal set.]

Generalizations of this work to a family of spatially homogeneous
anisotropic cosmologies will be presented elsewhere.

\nonumsection
Acknowledgments

We would like to thank John Friedman for discussions. This work was
supported in part by the NSF grants PHY90-05790 and PHY90-16733, and by
research funds provided by Syracuse University.


\nonumsection
References

\nref
R.~D. Sorkin,
talk given at the conference ``History of Modern Gauge Theories,''
held at Logan, Utah, July 1987, to appear in
{\it Int.\ J. Theor.\ Phys.\ \rm (1993).}

\nref
W.~G. Unruh,
\prd 40 1989 1048..

\nref
W.~G. Unruh and
R.~M. Wald,
\prd 40 1989 2598..

\nref
M.~Henneaux and C.~Teitelboim,
\plb 222 1989 195..

\nref
J.~D. Brown and
J.~W. York,
\prd 40 1989 3312..

\nref
R.~D. Sorkin,
in:
{\it Einstein Studies II,}
edited by A.~Ashtekar and J.~Stachel
(Birkhauser, Boston, 1991).

\nref
L.~Bombelli,
in: {\it Gravitation --- A Banff Summer Institute,}
edited by R.~Mann and P.~Wesson
(World Scientific, Singapore, 1991).

\nref
K.~V. Kucha\v{r},
\prd 43 1991 3332..

\nref
S.~W. Hawking, in: {\it Astrophysical Cosmology: Proceedings of the Study
Week on Cosmology and Fundamental Physics,} edited by H.~A. Br\"uck, G.~V.
Coyne and M.~S. Longair
(Pontificiae Academiae Scientiarum Scripta Varia,
Vatican City, 1982).

\nref
J.~B. Hartle and S.~W. Hawking,
\prd 28 1983 2960..

\nref
S.~W. Hawking,
\npb 239 1984 257..

\nref
J.~J. Halliwell and
J.~B. Hartle,
\prd 41 1990 1815..

\nref
J.~J. Halliwell and
J.~Louko,
\prd 42 1990 3997..

\nref
M.~Reed and B.~Simon,
{\it Methods of Modern Mathematical Physics\/}
(Academic, New York, 1980).

\nref
A.~Linde,
\zetf 87 1984 369.
[\sovpjetp 60 1984 211.];
\ncim 39 1984 401.;
\repprogph 47 1984 925..

\nref
A.~Vilenkin,
\prd 30 1984 509..

\nref
A.~Vilenkin,
\prd 33 1986 3560.;
\prd 37 1988 888..

\totalno

\bye